**Machine-Learned Interatomic Potential for Predictive Simulation of $MoS_2$ Epitaxy**

Emir Bilgili[1*], Nicholas Taormina[1], Richard Hennig[2], Simon R. Phillpot[2], Youping Chen[1]

[1] Department of Mechanical and Aerospace Engineering, University of Florida, Gainesville FL 32611

[2] Department of Materials Science and Engineering, University of Florida, Gainesville FL 32611

**Abstract**:

A machine-learned interatomic potential (MLIP) for multilayer $MoS_2$ was developed using the ultra-fast force field (UF3) framework. The UF3 MLIP reproduces key properties in strong agreement with DFT including lattice constants, interlayer binding energies, and phase-stability. Furthermore, the potential reasonably captures the phonon spectra and the highly anisotropic elastic tensor across monolayer (1H) and bulk (2H, 3R) $MoS_2$ phases. Critically, defect and edge formation energies are captured with high fidelity, exhibiting a strong correlation with DFT ($R^2$ = 0.91) across ten defective monolayers and reproducing the difference between the free energies of zigzag and armchair edges within 5% of DFT. Non-equilibrium molecular-dynamics simulations reveal layered homoepitaxial growth consistent with experimental observations, demonstrating the formation of van der Waals gaps between successive epilayers and triangular domains bounded by zigzag edges. The robust UF3 MLIP, which is only ~2X slower than the fastest empirical potentials, enables large-scale atomistic simulations of $MoS_2$ epitaxial growth.

** Corresponding author: emir.bilgili@ufl.edu*

## I. Introduction

Heterostructures incorporating single-and few-layer transition metal dichalcogenides (TMDs) hold promise for next-generation technologies including electronic, optoelectronic, memory, sensing, and energy/storage devices[1]. Molybdenum Disulfide ($MoS_2$) — with its sub-nm monolayer thickness, tunable bandgap, and thermal stability — is the most well-researched and promising TMD material, particularly for the development of post-silicon 2D transistors[2,3]. Though many-proof of concept $MoS_2$ heterostructures have been demonstrated, no commercial TMD device is currently available; a critical obstacle is the control and prediction of their manufacturing processes[4].

Epitaxy, the growth of thin films on crystalline substrates, remains the leading manufacturing method for the atomically sharp interfaces that solid-state devices require. Yet, the epitaxial growth dynamics of 2D semiconducting $MoS_2$ layers are not fully understood. Various mechanisms and governing principles regarding the growth dynamics of 2D materials have been proposed including symmetry-governed orientation selection, in which high-symmetry directions of the 2D lattice align with those of the substrate surface to maximize the symmetry of the combined system[5], edge-guided domain propagation[6,7], temperature-and-flux-driven orientation alignment[8–10], and defect-mediated pathways[11].

Molecular-dynamics (MD) simulations are a natural tool for probing these processes, but their success relies on the accuracy and efficiency of available interatomic potentials. Existing potentials for other material systems[12,13] have been successful in reproducing experimentally observed epitaxial growth dynamics, including defect formation and growth morphology[14]. However, developing high-fidelity interatomic potentials for *n*-layer $MoS_2$ is even more challenging because of its layered structure. This necessitates the concurrent treatment of strong covalent short-range in-plane bonding with weaker van der Waals (vdW) long-range interlayer bonding, a complexity for which even density functional theory (DFT) methods require explicit dispersion corrections to capture accurately[15–17]. In fact, most interatomic potentials available for $MoS_2$ focus exclusively on 2D monolayer $MoS_2$, thereby not having to deal with this significant anisotropy.

Classical empirical potentials for $MoS_2$ have been developed using a range of functional forms, including reactive bond-order (REBO)[18,19], Stillinger–Weber (SW)[20–23], and ReaxFF[24,25] formulations with various parameterization strategies such as valence force field fitting[21], particle-swarm optimization[22], and force-matching (FM) methods[23]. Although practical in many applications, existing empirical potentials were fit to a narrow set of structural or energetic benchmarks such as bond energies, bond lengths, bond angles, and intra- or inter-layer distances, as well as phonon-dispersion spectra or a subset of elastic properties. As such, they were not parameterized explicitly for and are unable to capture the non-equilibrium epitaxial processes of $MoS_2$ layers.

Machine-learned interatomic potentials (MLIPs) offer the prospect of near-DFT accuracy at the efficiency of MD. However, existing MLIPs for $MoS_2$ have largely prioritized minimizing force and energy errors near equilibrium or specific dynamic properties such as phonon dispersion relations and thermal transport or other specialized applications[26–36]. Like CEPs, many MLIPs focus exclusively on 2D $MoS_2$ and are not appropriate for multilayer systems. Furthermore, existing $MoS_2$ MLIPs remain orders of magnitude slower than CEPs, though much faster than DFT, limiting their accessible length and timescales required for epitaxial growth.

Recently, the Ultra-Fast Force Fields (UF3) framework, which combines effective two- and three-body potentials in a cubic B-spline basis with regularized linear regression has been developed. UF3 learns potentials that are physically interpretable and demonstrate accuracy as high as other advanced MLIP methods but can maintain speed similar to traditional empirical potentials[37]. This speed is essential for simulating highly nonequilibrium and size-dependent processes of epitaxy such as epilayer crystallization and defect formation, where large length and time scales must be accessed. Recently, UF3 has been successfully employed for the simulation of the epitaxial growth of AlN[38].

In this work, we develop an ultra-fast UF3 interatomic potential for semiconducting monolayer and multilayer $MoS_2$ that achieves near-DFT accuracy in structural prediction and interlayer binding, while reasonably reproducing the strongly anisotropic elastic tensor and phonon spectra. Critically, defect and edge energetics are consistent with DFT, preserving the formation energy hierarchy of various defects and the relative free energies of zigzag and armchair edges. Most importantly, the UF3 potential enables, for the first time, simulations of $MoS_2$ homoepitaxial growth consistent with experimental observations, demonstrating triangular domain formation, adatom incorporation along exposed zigzag edges, and stacking that preserves the characteristic van der Waals gap between monolayers.

By enabling the simulation of $MoS_2$ homoepitaxial growth, this work establishes the foundation for developing a $MoS_2$/substrate interatomic potential. The resulting $MoS_2$ MLIP represents a crucial first step toward predictive modeling of $MoS_2$/substrate heterostructures.

## II. Methodology

The development of the MLIP consisted of 3 steps: (1) dataset generation, (2) interaction-representation design and featurization, and (3) hyperparameter optimization and validation. The tool used for hyperparameter optimization was developed and reported in a previous paper[38].

### *2.1 Dataset generation*

Development of a machine-learned interatomic potential (MLIP) requires a sufficiently large and diverse dataset that spans the relevant domains of the potential energy surface (PES). For the UF3 framework, each training entry must include atomic coordinates, total energy, and atomic forces. The dataset must adequately sample atomic environments relevant to the target application while maintaining enough diversity to ensure generalizability. This diversity prevents overfitting to specific configurations and enables the potential to accurately predict unseen atomic arrangements encountered during atomistic simulations.

To rapidly generate a training set, the Genetic Algorithm for Structure and Phase Prediction (GASP) was employed[39,40]. GASP uses evolutionary algorithms to efficiently sample composition spaces, generating the structure and relaxation trajectories of thousands of stable and metastable configurations in DFT (Fig. 1). GASP has been successfully employed to develop UF3 MLIPs for other material systems[38,41] including for the simulation of surface processes and homoepitaxy.

In this work, the following types of data were included to comprehensively represent the Mo–S system: (1) both stoichiometric ($MoS_2$) and nonstoichiometric (across Mo-S composition space) GASP searches, (2) fully relaxed polymorphs from Materials Project[42], (3) structures relaxed under finite external pressure, (4) elastically strained structures via VASPKIT[43], (5) lattice-distorted structures via Phonopy[44,45], (6) ab-initio molecular-dynamics (aiMD) trajectories at 600 K, 1500 K, and 3000 K for up to 5 ps, (7) slabs and surfaces, (8) isolated monolayers and nanoribbons.

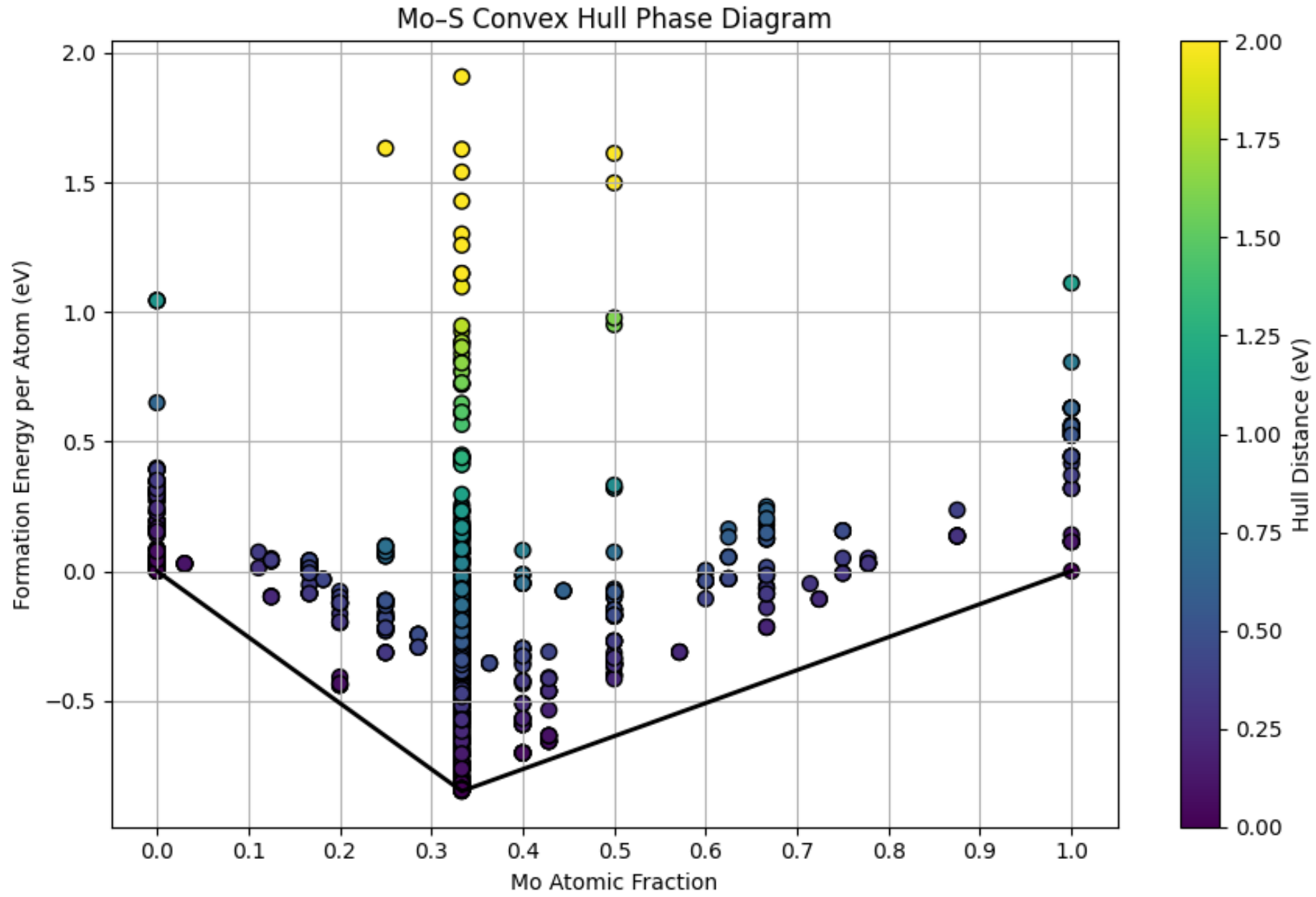

**Fig. 1. GASP identified structures within the Mo-S composition space.** A structure's energy above hull is indicated by the intensity of its circular marker. Only relaxed structures are displayed.

DFT calculations were performed using the projector augmented-wave (PAW) method[46] as implemented in Vienna Ab Initio Simulation Package (VASP)[47,48]. The exchange–correlation energy was described using the Perdew–Burke–Ernzerhof (PBE) functional[49]. A plane-wave cutoff of 520 eV was used, and vdW interactions were explicitly included using the DFT-D3 correction of Grimme[50]. Brillouin zone integrations employed Gaussian smearing with a width of 0.05 eV. For bulk structures, Γ-centered meshes with a density of 40 $Å^{-1}$ were used; for slabs and monolayers, Γ-centered 6×6×1 Monkhorst–Pack grids were adopted; and for nanoribbons, a Γ-centered 12×1×1 grid was employed. All slab and nanoribbon models used 20 Å of vacuum and applied dipole corrections. Structural relaxations were converged to $10^{-6}$ eV/atom in electronic energy and $10^{-5}$ eV/Å in ionic forces.

Prior to featurization, all structures were rigorously filtered to ensure physically meaningful data. Configurations with any force component exceeding ±20 eV/Å or atomic separations shorter than 0.5 Å were discarded as not being physically realistic. Energy-per-atom (EPA) windows of –7.55 to –5.0 eV/atom were applied to the stoichiometric GASP, pressure, aiMD, surface, and monolayer datasets to remove spurious low-energy entries arising from corruption and exclude highly unstable configurations. A broader EPA window of –12.0 to –3.8 eV/atom was applied to the GASP dataset across composition space to preserve chemically diverse, non-stoichiometric structures. To reduce redundancy, farthest-point sampling (FPS) was applied in EPA space, enforcing a minimum separation of $4×10^{-4}$ eV/atom between structures in the GASP, pressure, and aiMD datasets.

Before filtering the dataset consisted of 33,614 structures. Following consolidation across all sources, the final dataset comprised 28,579 structures, partitioned into 25,792 for training and 2,787 for testing. This dataset provided broad coverage of both equilibrium and non-equilibrium configurations.

### *2.2 Interaction representation design*

The UF3 framework represents atomic interactions through cubic B-spline interpolations of two- and three-body terms. There are two fundamental design choices: the cutoff distances (minimum and maximum radii for 2- and 3-body terms) and the knot mapping (spline spacing and density). The cutoff distances directly impact the speed of the resulting interatomic potential as they control the number of neighbor interactions considered during simulation, with the 3-body cutoffs being the dominant factor. The knot mapping controls the computational demands of the featurization and fitting process as it determines the number of basis functions and dimensionality of feature space. The fidelity of the potential depends on both choices. Chosen cutoffs must be able to capture the relevant physical interactions without introducing unnecessary computational overhead at run-time. Likewise, knot resolution must be fine enough to represent sufficient details in the interaction landscape but not so dense that the model overfits to the training data or becomes computationally intractable during fitting.

For the two-body terms, a cutoff range of 0.001–9 Å with 54 internal uniform knots was employed. The extended cutoff ensures inclusion of long-range van der Waals interactions necessary to stabilize the layered $MoS_2$ structure, while the knot density provides adequate flexibility to resolve short- and medium-range covalent interactions. The final knot spacing was determined through heuristic testing and physical intuition to achieve a balance between representational accuracy and fitting efficiency. We note that shorter cut-offs (5-8 Å) were unable to accurately capture the vdW gap. For the three-body terms, the two short legs were defined from 0.5 to 3.6 Å with 10 internal knots and the long leg necessarily extended to 7.2 Å, with 20 internal knots. The 3-body cutoffs encompass all first-neighbor in-plane interactions (Mo–S ≈ 2.4 Å, Mo–Mo/S–S ≈ 3.2 Å) and associated angular correlations within the trigonal prismatic coordination. Together, these cutoff and knot configurations provide a balance between spatial resolution and computational efficiency, enabling accurate representation of both covalent and van der Waals interactions in $MoS_2$.

Featurization was automatically performed by UF3Tools[38] which imports and labels/splits the DFT dataset and uses of UF3's native environmental atomic descriptor. UF3 encodes invariant transformation of atomic coordinates using the *many-body expansion*[51], truncated at the 3-body term. Each term is expressed as a function of pairwise distances (one distance for two-body and three-distances for the three-body term). The 2B and 3B terms are represented as a linear combination and tensor product of cubic B-splines, respectively. The cubic B-spines are constructed with compact support (meaning that they are only non-zero over 4 adjacent intervals), such that the evaluation of any 2B and 3B potential requires the calculation of at-most 4 and $4^3$=64 basis functions, respectively[37]. This construction is what makes UF3 computationally efficient at run-time and on-par with the speed of empirical potentials (see Section 3.6). For a given cut-off radius (assuming uniform knot spacing), the speed of a UF3 potential is fixed, regardless of the shape of the potential function learned.

### *2.3 Hyperparameter optimization and validation*

Over 7,000 candidate UF3 potentials were rapidly fit using the UF3Tools software package, which has been recently developed[38]. UF3Tools uses Optuna's Tree-structured Parzen Estimator (TPE) algorithm[52] to perform a global search over key hyperparameters, including force-to-energy weights, regularization strengths, curvature penalties, and dataset sample weights for multiple $MoS_2$ phases (1H, 2H, 3R). UF3Tools automates both training and validation by testing each potential in atomistic simulations (LAMMPS[53,54]) and comparing predicted properties against corresponding reference values (DFT). The objective function in UF3Tools normalizes each property's absolute error by a target tolerance (e.g., 5% for lattice constants, 10% for elastic constant $C_{11}$, 15% for elastic constant $C_{13}$), ensuring that deviations in one metric do not dominate the optimization. Metrics considered in optimization include root-mean squared error (RMSE) in energy and force, lattice constants, elastic constants, and surface energies. For further and detailed methodological information on UF3 tools, the reader is referred to the original paper[38]. Minor modifications were made to the tool to account for the validation of 2D material monolayers which is maintained on the UF3Tools GitHub repository. Following the initial structural and mechanical validation, the models were further evaluated across three successive validation stages to ensure transferability and physical fidelity.

The next validation stage comprised two subphases aimed at assessing the vibrational fidelity and thermal stability of the candidate potentials. In the first subphase, phonon band dispersions and densities of states were computed using Phonolammps[55] for the UF3–LAMMPS simulations benchmarked against DFT calculations using Phonopy[44,45]. The resulting phonon spectra showed an acceptable root-mean-square (RMS) deviation of approximately 0.6 THz across the dispersion curves, with qualitative agreement across acoustic and optical branches. The basis for the performance of phonon spectra deemed acceptable is consistent with typical RMS errors of ~0.5 – 1 THz commonly reported in MLIPs[56]. Second, finite-temperature molecular-dynamics simulations were performed on slab systems heated up to 1500 K to confirm dynamical stability and preservation of the interlayer van der Waals gap. The next validation stage assessed the defect and edge energetics of 1H $MoS_2$ monolayers. Defects and edges play a critical role in epitaxial processes of 2D materials, acting as nucleation and adatoms incorporation sites. Finally, models were used to simulate epitaxial growth, with potential performance assessed by the potentials ability to capture the crystallization process and maintain the van der Waals gap between the growing layer and the substrate.

## III. Results and Discussion

In this section we outline the performance of the developed $MoS_2$ interatomic potential and draw some comparisons to existing potentials in the literature. We also demonstrate the novel capabilities of the developed potential in the non-equilibrium molecular-dynamics simulation of homoepitaxial growth of $MoS_2$ layers.

### *3.1 Selection of the final model*

Among the tested candidates, the final UF3 potential was selected based on its combined accuracy and stability across all validation stages. It simultaneously sufficiently reproduced:

- Lattice constants and interlayer binding energies in strong agreement with DFT, while maintaining reasonable elastic behavior and capturing the relative stability between phases;
- Phonon spectra and finite-temperature stability of multilayer substrates;
- Relative formation energies of defect with correlation $R^2 > 0.85$ and the ratio between the free energies of zigzag and armchair edges below 10% error;
- Stable epitaxial growth, capturing epilayer crystallization and maintaining the van der Waals gap between layers;

Among the tested candidates, the final UF3 potential was selected because it was the only model that remained accurate and stable across *all* validation stages. More than 85% potentials generated by UF3Tools reproduced lattice constants within 5% of DFT, and roughly half maintained the correct phase-stability ordering. Requiring an average surface-energy error near 10% across 2H and 3R stackings reduced the pool to ~600 potentials. Enforcing ordering and accuracy of elastic constants further narrowed the candidates to ~3% of the global population (~250 potentials). Phonon spectra and finite-temperature slab stability filtered the set to ~50. Only 36 potentials were deemed sufficient to attempt epitaxial growth simulations, and **one** of these successfully reproduced epilayer crystallization, maintained the van der Waals gap, and remained dynamically stable during adatom deposition. This final model was therefore selected as the UF3 potential for subsequent production runs of epitaxial simulations. Many potentials with adequate or even superior (to the final potential) static and dynamic material properties were unable to simulate epitaxial growth, demonstrating either (1) unstable behavior during deposition and/or incorporation of adatoms at the substrate surface or (2) unphysical epilayer morphologies (e.g., bunching up into 3D clusters or not preserving the vdW gap). The UF3-$MoS_2$ GitHub associated with this work provides the UF3 potential file.

Figure 2 presents the energy and force root-mean-square errors (RMSE) of the selected UF3 MLIP. While the fitted model yields RMSE values of 0.117 eV for energies and 0.420 eV Å$^{-1}$ for forces, minimizing these numerical metrics is not the primary objective. In practice, such differences in RMSE had little influence or predictability on the potential's ability to simulate epitaxial growth. This is because the reference dataset spans an exceptionally broad range of atomic configurations to capture highly nonequilibrium environments — an essential requirement for epitaxial growth simulations. For example, the per-atom energies range from −7.55 to −5.5 eV, and atomic forces extend between −15 and +15 eV Å$^{-1}$, nearly an order of magnitude larger than the ranges typically used in MLIP development for near-equilibrium $MoS_2$. As a result, to be able to simulate non-equilibrium epitaxial growth, the UF3 potential must remain transferable across both low-energy crystalline and high-energy disordered states rather than be narrowly optimized around equilibrium. Consequently, while this broader configuration space naturally increases the numerical RMSE, the model's fidelity in reproducing key physical properties—such as structural stability, phonon dispersions, defect and edge energetics—demonstrates that it captures the underlying physics. Accordingly, the UF3 model was optimized to preserve these physical properties rather than to minimize absolute energy and force errors.

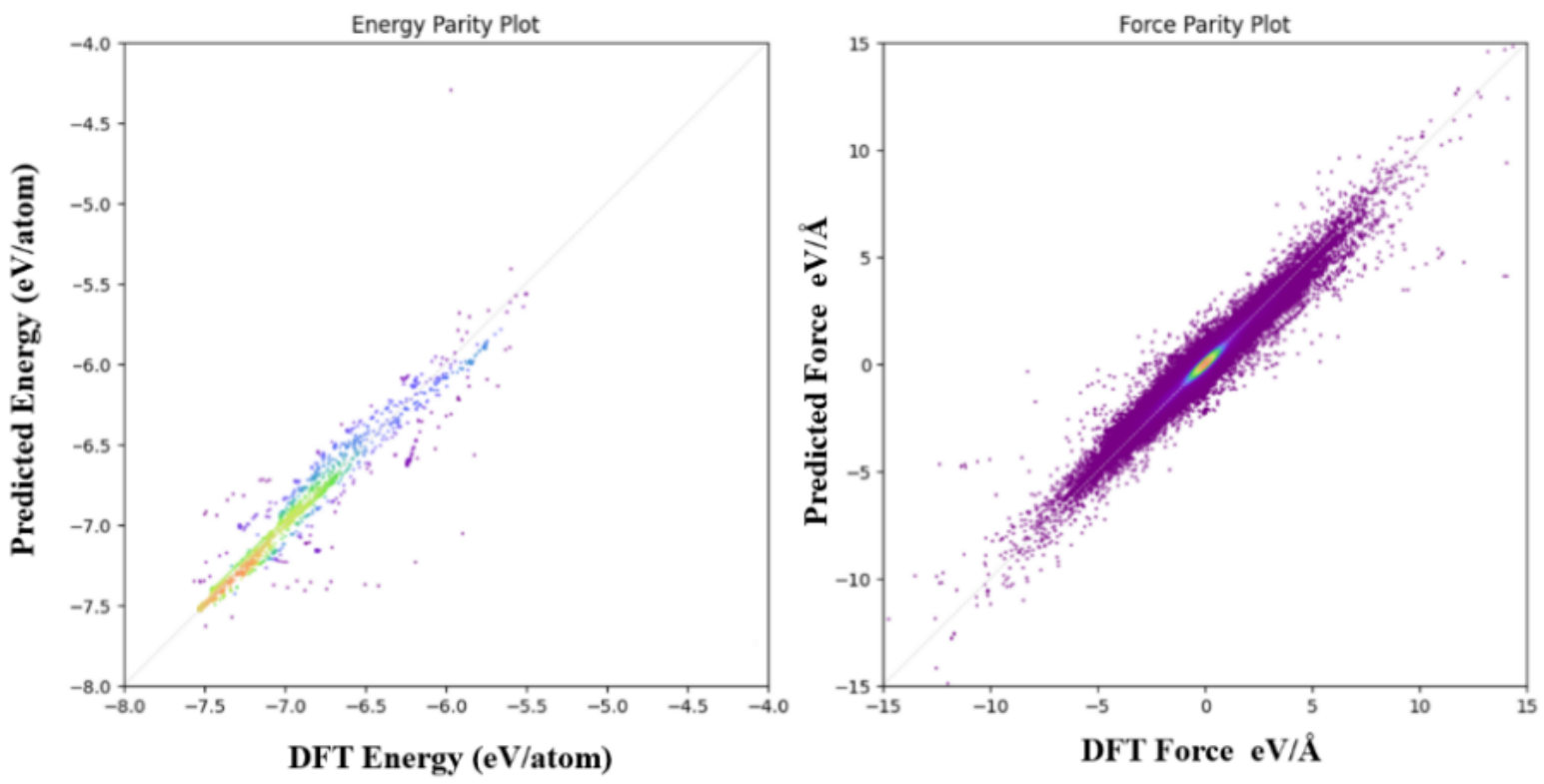

**Fig. 2. Energy (left) and force (right) parity plots of the testing dataset during potential fitting.** The model achieves an RMSE of 0.117 eV (MAE = 0.064 eV) for energies and 0.420 eV/Å (MAE = 0.137 eV/Å) for forces.

Figure 3 shows the learned two-body (2B) potentials for S–S, S–Mo, and Mo–Mo interactions. The very short-range regions (below ~1.6 Å, corresponding to pair distances unincluded in the filtered dataset) exhibit *spurious wells* within the nominally monoatomic repulsive regime. These features are numerical artifacts arising from spline interpolation in regions where no training data were available—since such unphysically short pair distances never occurred in the DFT reference set. Consequently, the splines are unconstrained in these intervals and can introduce non-monotonic behavior. Importantly, these artifacts do not affect the stability or accuracy of the simulations, as atoms in $MoS_2$ do not physically sample these very short-range configurations during equilibrium or non-equilibrium dynamics. The physically relevant portion of each potential (near and beyond the first-neighbor distance) is smooth and consistent with expected bonding behavior, ensuring stable dynamics. The S–S and Mo–Mo potentials exhibit deep short-range wells corresponding to their single-crystal lattice spacings followed by strong repulsive shoulders into attractive wells after 3 Å consistent with their configuration in hexagonal $MoS_2$ monolayers. The S–Mo potential displays a single pronounced well near the Mo–S bond length within monolayer $MoS_2$, followed by a strong repulsive shoulder around 3 Å that enforces the van der Waals gap between adjacent layers. Notably, at distances above 4 Å, the 2B potentials exhibit a complex energetic landscape with various local minima and repulsive regions, highlighting the models learning of interlayer and longe-range behavior.

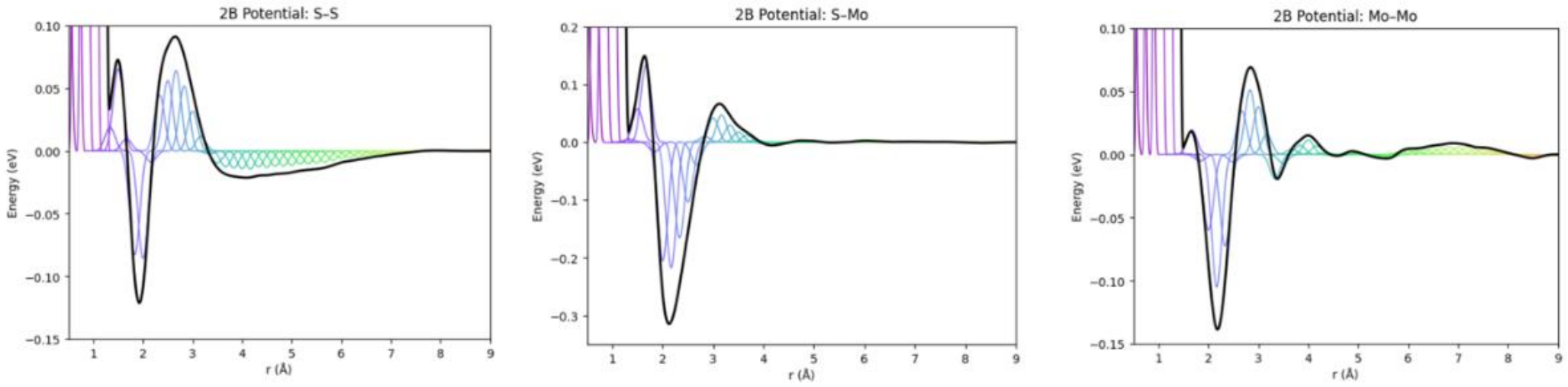

**Fig. 3. Learned two-body interaction potentials for Mo–Mo, S–S, and Mo–S pairs in the UF3 model.** Black lines show the fitted potentials, while colored curves indicate the underlying B-spline basis functions.

The short-range, intralayer interactions—responsible for preserving the trigonal-prismatic coordination—are primarily governed by the three-body term, which enforces the bond-angle preferences. The three-body surfaces, shown in Fig. 4., exhibit bean-shaped topologies, with sharp repulsive regions away from the ideal trigonal-prismatic geometry with penalties above 0.4 eV and neutral or mildly attractive pockets corresponding to unpenalized angular configurations. In this sense, the two-body terms set near-equilibrium nearest neighbor distances and long-range interlayer behavior, while the three-body term dominates the short-range energetics by strongly penalizing angular distortions that deviate from the trigonal-prismatic coordination, thereby maintaining the structural anisotropy of $MoS_2$. Notably, within the design constraints imposed by the chosen cutoffs and knot scheme, the UF3 MLIP *learned* this division of labor strategy—assigning much of the short-range, angle-dependent behavior to the three-body terms and the bulk of the long-range interlayer physics to the two-body terms.

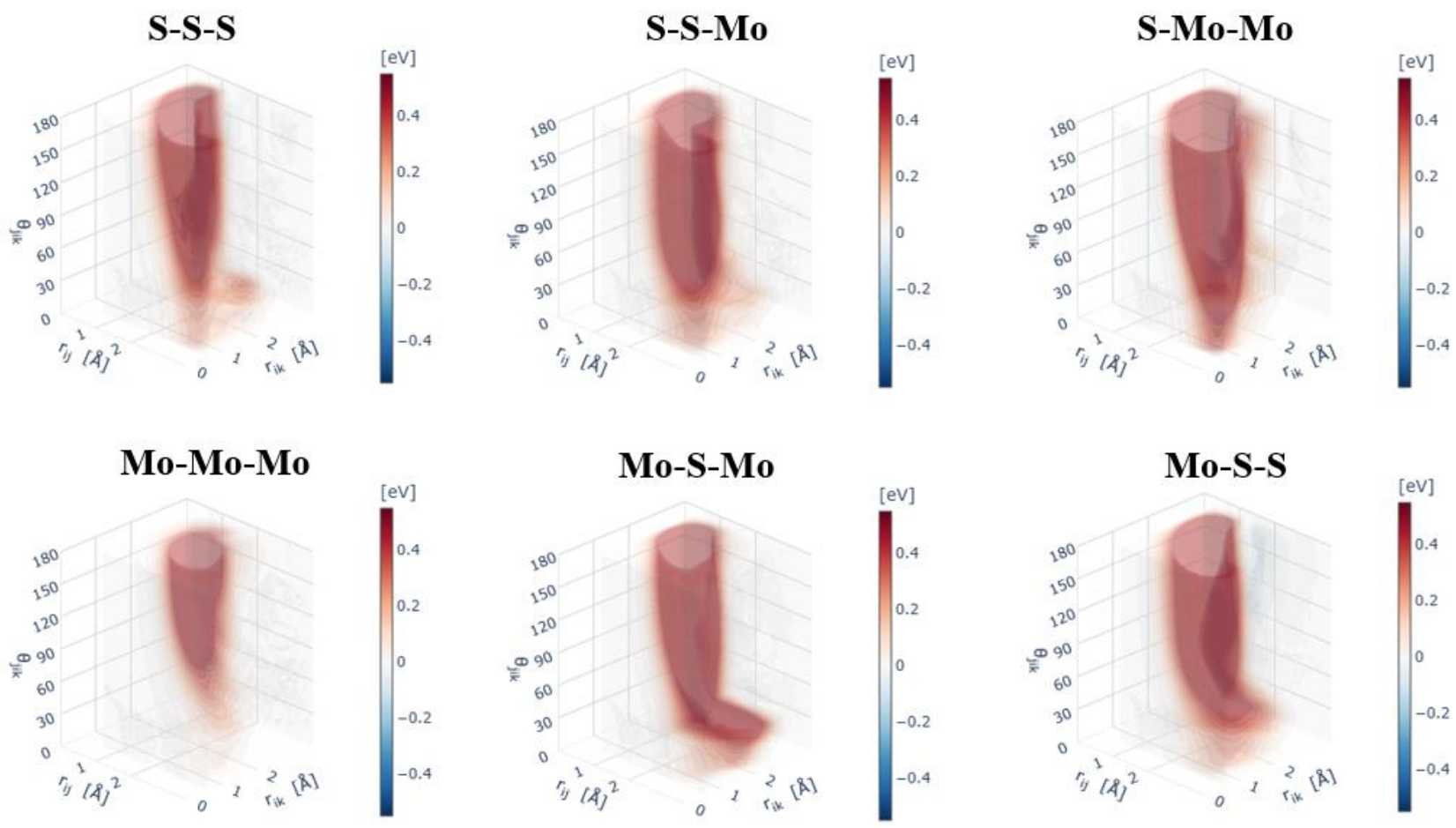

**Fig 4. Learned three-body (3B) interaction potentials for unique trios in the UF3 model.** Each surface is plotted as a function of the two radial legs $r_{ij}$, $r_{ik}$, and the bond angle $\theta_{ijk}$. The 3B surfaces exhibit bean-shaped topologies with sharp repulsive regions (penalties > 0.4 eV) away from the ideal trigonal-prismatic geometry and neutral or mildly attractive pockets near equilibrium. These topologies reflect the model's learned strategy of assigning short-range, angle-dependent intralayer penalties to the 3B terms, while the 2B terms capture the near-equilibrium nearest neighbor distances and weaker long-range interlayer behavior characteristic of anisotropic layered $MoS_2$.

### *3.2 Structural Parameters, Energetic Relationships, and Interlayer Binding*

**Table I** summarizes the equilibrium structural parameters, cohesive energies, and surface (interlayer binding) energies of the multilayer (2H, 3R) and monolayer (1H, 1T′) $MoS_2$ polytypes. For monolayer phases, $t$ denotes the intralayer S–S separation sandwiching the Mo atomic plane, serving as a measure of the sheet thickness.

UF3 reproduces the lattice parameters of the semiconducting 1H, 2H, and 3R $MoS_2$ phases with DFT-level accuracy (< 2 % error), and reproduces the interlayer van der Waals gap in bulk 2H and 3R $MoS_2$ with errors of ~1% and ~3%, respectively (2H: DFT = 3.0343 Å, UF3 = 3.0662 Å, 3R: DFT = 3.0266 Å, UF3 = 3.1277 Å). UF3 also demonstrates some transferability to the metallic 1T′ phase, yielding < 4 % deviation in in-plane lattice constants and an ~ 8 % underestimation of sheet thickness. Although the cohesive energies are systematically offset by 66 – 71 % (DFT ~ -5 eV/atom, UF3 ~ -2 eV/atom), UF3 accurately captures the relative energetic ordering among the polymorphs (2H<3R<1H<1T'), a critical requirement for modeling phase stability, transformation, and crystal growth. The inability of the UF3 potential to capture the absolute values of the cohesive energy is likely due to the prioritization of other parameters during hyperparameter optimization and may also be due to the aforementioned dominance of the 3B term in describing intralayer cohesion (which is the dominant energetic contribution to the total cohesive energy).

Fitting energy–volume data to a third-order Birch–Murnaghan equation of state[57] yield bulk moduli $K_0$ of 81.7 (72.1) GPa for 2H-$MoS_2$ and 77.4 (72.8) GPa for 3R-$MoS_2$, with corresponding pressure derivatives $K_0'$ of 4.50 (4.11) and 3.89 (4.10), respectively (DFT values in parentheses). The UF3 potential closely reproduces the DFT response under homogeneous compression, while exhibiting a nearly uniform over-stiffening in the tensile regime. We note that the equation-of-state analysis intentionally spans a wide volumetric strain range of ~ $0.68 < V/V_0 < 1.4$, as, in this work, we are concerned with the development of an interatomic potential with high fidelity far from equilibrium. The Supplemental Material (SM) associated with this work provides the energy-volume plots of UF3 vs. DFT (Fig. S1 and Table SI).

**Table I. Structural parameters, cohesive energies, and interlayer binding energies of $MoS_2$ polytypes from DFT and UF3**

| Phase | DFT | | | | | | UF3 | | | | | |
|---|---|---|---|---|---|---|---|---|---|---|---|---|
| | a (Å) | b (Å) | c (Å) | t (Å) | $E_{coh}$ (eV/atom) | $\gamma_{0001}$ (eV/nm$^2$) | a (%$^{err}$) | b (%$^{err}$) | c (%$^{err}$) | t (%$^{err}$) | $E_{coh}$ eV/atom [%$^{err}$] | $\gamma_{0001}$ (%$^{err}$) |
| 2H | 3.1615 | 3.1615 | 12.3215 | - | -5.2630 | 1.5174 | -0.12 | -0.12 | 0.60 | - | -1.80 [-65.80] | 2.30 |
| 3R | 3.1650 | 3.1650 | 18.4498 | - | -5.2627 | 1.5236 | -0.41 | -0.39 | 1.74 | - | -1.79 [-65.90] | 18.02 |
| 1H | 3.1615 | 3.1615 | - | 3.1358 | -5.1792 | - | -0.20 | -0.20 | - | 0.11 | 1.68 [-67.50] | - |
| 1T′ | 5.6925 | 3.1607 | - | 3.4702 | -4.9972 | - | -3.41 | 0.43 | | -8.04 | 1.45 [-71.08] | - |

***For bulk phases (2H, 3R), c is the out-of-plane lattice constant. For monolayer phases (1H, 1T'), t denotes the intralayer S–S separation sandwiching the Mo atomic row. For cohesive energies, $E_{coh}$, both the raw value in eV/atom and percent error (in brackets) is given on the right-hand side of the table so as to easily discern reproduction of the relative stability between phases.***

To determine the developed potential's ability to capture the interlayer binding and strength of the long-range vdW forces, the surface energy of the basal plane (0001) was computed. UF3 reproduces the basal plane surface energy of multilayer $MoS_2$ with high fidelity, achieving 2.3 % and 18 % error for the 2H and 3R phases, respectively. Further, it preservers the energetic ordering of the surfaces: $\gamma_{DFT}^{3R} = 1.523 \frac{eV}{nm^2} > \gamma_{DFT}^{2H} = 1.517 \frac{eV}{nm^2}$ ; $\gamma_{UF3}^{3R} = 1.798 \frac{eV}{nm^2} > \gamma_{DFT}^{2H} = 1.555 \frac{eV}{nm^2}$. Accurate description of interlayer interactions is crucial for exfoliation, stacking, and epitaxial growth modeling. Although UF3 underestimates the absolute cohesive energies, this is not inconsistent with its accurate prediction of surface energy. In DFT, the cohesive energy of bulk 2H-$MoS_2$ (~ −5.26 eV/atom) differs from that of an isolated monolayer (~ −5.18 eV/atom) by only ~ 0.08 eV/ atom, less than 2 % of the total binding. Thus, even if the potential uniformly underestimates the total cohesive energy (dominated by strong intralayer covalent bonds), it can still accurately capture the much weaker out-of-plane vdW bonding that determines the basal-plane surface energy.

Figure 5 compares the percent error relative to DFT in predicted basal-plane (0001) surface energies of 2H- and 3R-$MoS_2$ across various existing potentials. Nearly all conventional models give significantly lower values, with deviations ranging from ~ 20 % to > 90 %. This is expected, as most were parameterized exclusively for monolayer $MoS_2$; by design, therefore, they lack explicit fitting to interlayer or adhesion data. REBO[19], SW-Jiang[20], and SNAP[26], in particular, predict binding energies nearly an order of magnitude too small. To compare the performance of UF3 to universal MLIPs, we include two representative off-the-shelf models: MACE-MP-0[58,59] (v0.3.14) and CHGNet[60] (v0.3.0). For both 2H- and 3R-$MoS_2$, these universal MLIPs predict negligible basal-plane surface energies of ~0.022 eV $nm^{-2}$ (MACE-MP-0) and 0 eV $nm^{-2}$ (CHGNet), corresponding to errors of -98.5% and -100%, respectively. This is likely due to the underlying Materials Project Trajectory (MPtrj) dataset[42,59,60] used to develop these universal potentials which (1) has a limited representation of surface structures and primarily consists of bulk crystals, and (2) lacks explicit long-range dispersion corrections (e.g., DFT-D3). By contrast, ReaxFF[25] yields moderate agreement with DFT, consistent with its inclusion of adhesion energies during fitting. Somewhat surprisingly, the SW-FM[23] potential also performs reasonably well despite being developed solely for the monolayer, suggesting that its functional form or parametrization preserved sufficient nonbonded interaction strength to capture aspects of interlayer cohesion. Among compared models, UF3 provides the best overall agreement, achieving quantitative accuracy for the 2H phase (2 % error) and a manageable 18 % overestimation for 3R.

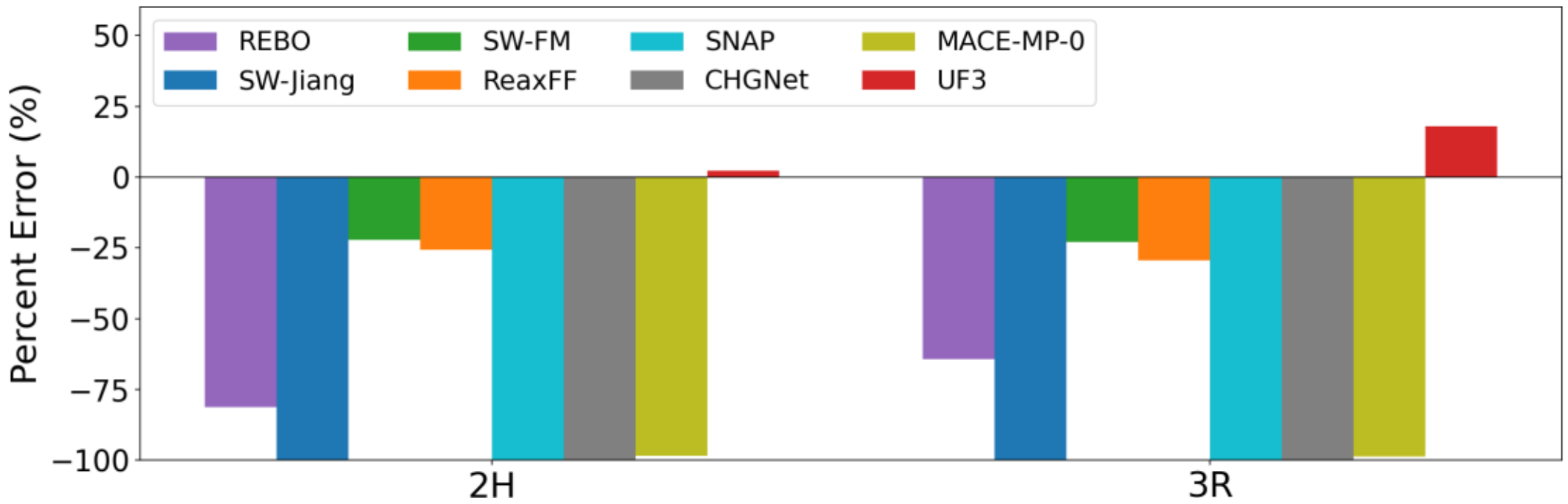

**Fig. 5. Percent errors in (0001) surface energy predictions relative to DFT for 2H- and 3R-$MoS_2$ across various interatomic potentials.** Existing potentials systematically underestimate interlayer binding, whereas UF3 overestimates by only 3% (2H) and 18% (3R).

### *3.3 Phonon Spectra and Elastic properties*

Figure 6 presents the phonon band structures for semiconducting $MoS_2$ phases. UF3 reproduces the acoustic branches with good fidelity, showing small bias below 7 THz (mean deviations of –0.02, –0.14, and –0.03 THz for 1H, 2H, and 3R, respectively). This ensures that elastic and second-order acoustic properties are well described. In contrast, the optical branches above 7 THz are systematically softened relative to DFT, with mean deviations of –0.46, –0.47, and –0.34 THz, respectively. These shifts result in modest downward displacements of the high-frequency density of states (DOS) peaks. Quantitatively, UF3 yields overall band RMS errors of 0.49, 0.52, and 0.39 THz for 1H, 2H, and 3R, with DOS RMS error of 0.87, 0.83, and 0.67. While high-frequency optical modes are not reproduced very accurately, the errors remain modest and are competitive with existing potentials. In semiconducting $MoS_2$, the thermal conductivity is dominated by acoustic phonons, which possess much larger group velocities and significantly longer mean free paths than the flat, low-velocity optical branches[61]. As such, the larger errors in the optical modes should have only a limited impact on thermal transport behavior.

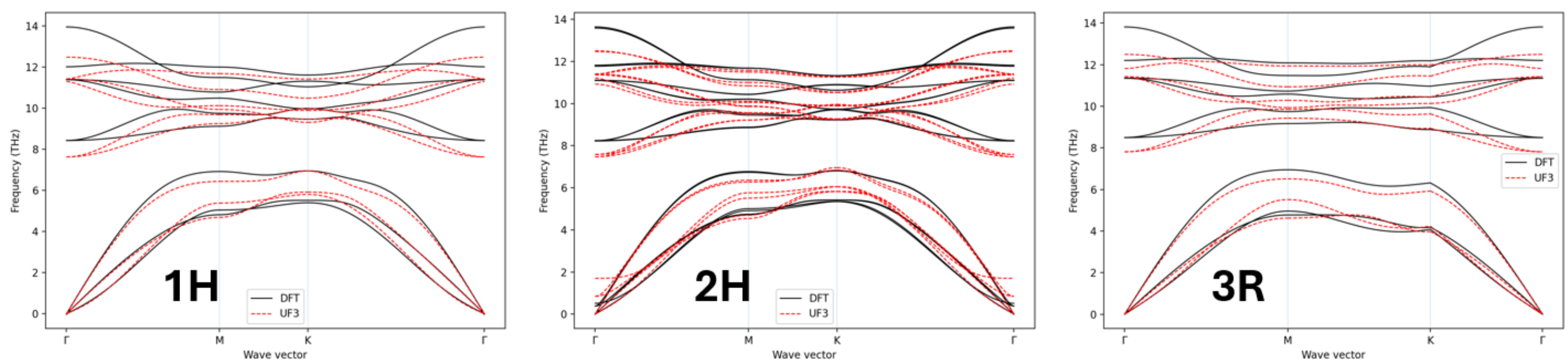

**Fig. 6 . Phonon band structures for semiconducting monolayer (1H) and bulk (2H, 3R) $MoS_2$.** UF3 predictions (red dashed lines) are compared against DFT reference calculations (black lines).

Accurately reproducing the elastic tensor of $MoS_2$ is significantly challenging due to its strong anisotropy, where in-plane elastic components ($C_{11} \sim 220\ GPa$) exceed out-of-plane components (e.g., $C_{13} \sim 10\ GPa$ , $C_{33} \sim 45\ GPa$) by an order of magnitude. Indeed, the majority of existing MLIPs developed for $MoS_2$ omit explicit reporting of elastic constants[28,30–33,35] or reduce validation to simplified metrics or proxies such as Young's modulus of the monolayer[30], energy–strain curves[29], or a small subset of constants[34]. Similarly, many existing classical potentials only report a subset of elastic constants (e.g., $C_{11}$ and $C_{33}$ only)[18,19], if at all[25], or have focused exclusively on monolayer $MoS_2$[20,21,23]. **Table II** compares the predicted elastic constants of 2H-$MoS_2$ and 3R-$MoS_2$ of the UF3 potential against DFT. In reporting, we adopt the standard Voigt notation in which indices 1–6 correspond to $xx$, $yy$, $zz$, $yz$, $xz$, and $xy$, respectively. Overall, UF3 achieves a physically reasonable representation of $MoS_2$ elasticity, preserving both the ordering and magnitude hierarchy of elastic constants, namely: $C_{11} > C_{12} > C_{33} > C_{44} \sim C_{13} > C_{14}$.

2H-$MoS_2$ possesses five independent elastic constants ($C_{11}$, $C_{12}$, $C_{13}$, $C_{33}$, and $C_{44}$) with $C_{66} = \frac{1}{2}(C_{11}-C_{12})$. In contrast, the rhombohedral symmetry of the 3R phase admits an additional independent constant, allowing a nonzero $C_{14}$. UF3 correctly captures these symmetry distinctions, predicting $C_{14} = 0$ for 2H-$MoS_2$ and a small finite value for 3R-$MoS_2$, in agreement with DFT.

**Table II. Predicted Elastic Constants of 2H and 3R $MoS_2$**

| Elastic Constant (GPa) | 2H | | 3R | |
|---|---|---|---|---|
| | DFT | UF3 | DFT | UF3 |
| $C_{11}$ | 222.1 | 227.1 | 222.9 | 267.9 |
| $C_{12}$ | 52.6 | 81.0 | 54.2 | 101.7 |
| $C_{13}$ | 9.2 | 7.4 | 12.2 | 24.6 |
| $C_{14}$ | - | 0.00 | 3.7 | 2.3 |
| $C_{33}$ | 46.2 | 71.2 | 42.5 | 63.0 |
| $C_{44}$ | 14.2 | 15.6 | 15.3 | 24.2 |
| $C_{66}$= ½ ($C_{11}$-$C_{12}$) | 84.7 | 73.0 | 84.4 | 83.1 |

The UF3 potential overestimates the so-called 2D "elastic constants" of isolated $MoS_2$ monolayers (UF3: $C^{2D}_{11}$ = 164 N/m, $C^{2D}_{12}$ = 59 N/m; DFT: $C^{2D}_{11}$ = 132 N/m, $C^{2D}_{12}$ = 32 N/m), a consequence of the longer-range interactions needed to capture interlayer van der Waals cohesion, which also increase intralayer binding. However, since the elastic constant offset is systematic, the in-plane shear response $C^{2D}_{66}$ =½ ($C^{2D}_{11}$– $C^{2D}_{12}$) = 55 N/m is within ~2% of DFT.

The 2D "elastic constants" for monolayer film with vacuum spacing is defined as:

$$C_{ij}^{2D} = C_{ij}^{3D} \times L_z,$$

where $C_{ij}^{3D}$ are the conventional 3D elastic constants (in GPa) and $L_z$ is the supercell height[62,63]. This conversion yields thickness-independent in-plane "elastic constants" with units of N/m. The underlying assumption is that the S–Mo–S triatomic monolayer behaves intrinsically as a two-dimensional elastic sheet, with no well-defined out-of-plane thickness. These 2D elastic constants are commonly reported for 2D material monolayers, including in the interatomic potential performance of $MoS_2$[64] .

Despite some quantitative limitations in elasticity, UF3 captures the low-frequency acoustic phonon modes of monolayer (1H) $MoS_2$ with high fidelity. These modes correspond to the long-wavelength limit of the lattice dynamical response ($q \rightarrow 0$), directly tied to the elastic-wave velocities. In homoepitaxy, where no lattice mismatch strain exists, inaccuracies in elastic tensor behavior are less consequential, but vibrational fidelity is essential for realistic finite-temperature dynamics. Therefore, UF3 prioritized accurately capturing monolayer vibrational spectra while balancing elasticity across 1H, 2H, and 3R phases.

### *3.4 Defect and Edge Energetics*

To quantify the ability of the UF3 potential in capturing $MoS_2$ defect behavior, relative defect formation energies were computed using a 5×5 (75-atom) pristine monolayer as the reference configuration, $\Delta E^{\%}_{\text{defect}}$, defined as

$$\Delta E^{\%}_{\text{defect}} = \frac{E_{\text{defect}} - E_{\text{pristine}}}{|E_{\text{pristine}}|} \times 100.$$

Where $E_{defect} - E_{\text{pristine}} = \Delta\text{E}_{\text{defect}}$ is the energy difference between the defective and pristine monolayer. Because UF3 exhibits a systematic cohesive-energy offset relative to DFT, relative defect formation energies were used instead of absolute values to ensure an apples-to-apples comparison. Normalizing by the energy of pristine reference cancels this offset and ensures the comparison is uncorrupted by errors in cohesive energy. The Supplemental Material provides raw energy values predicted by both DFT and UF3 for the defect structures considered in the following comparison (see Table SII).

The considered defects include commonly observed configurations during epitaxial growth — $V_S$ (single sulfur vacancy), $V_{S_2}$ (disulfur vacancy), $V_{Mo}$ (molybdenum vacancy), $Mo_{S_2}$ (antisite defect where an Mo atom substitutes an $S_2$ column), $S_{2Mo}$ (antisite defect where an $S_2$ column substitutes an Mo atom), $V_{MoS_3}$ (Mo vacancy with three neighboring S atoms removed), and $V_{MoS_6}$ (Mo vacancy with all six neighboring S atoms removed)—as well as additional defects such as $V_SV_S$ (a sulfur divacancy consisting of one S atom removed from the top sublayer and one from the bottom sublayer with different in-plane coordinates) and a Schottky-like defect ($\text{Schottky}_{\text{vS}_2}$) consisting of a simultaneous Mo vacancy ($V_{Mo}$) and disulfur vacancy ($V_{S_2}$). Together, these defects capture both experimentally observed configurations and hypothetical vacancy complexes used for benchmarking. Figure 7 compares relative defect formation energies for a 1H-$MoS_2$ monolayer, computed with UF3 and DFT, with all values referenced to relaxed defective structures relative to a pristine 75-atom 5×5 supercell. The inset images show atomically resolved experimental ADF micrographs of representative defects in monolayer $MoS_2$[65]. UF3 captures the qualitative hierarchy of defect energies reasonably well, particularly for experimentally observed defects. We reiterate that since UF3 systematically underestimates $MoS_2$ cohesive energies, we considered only relative defect energies which were expressed as the percentage of energy lost relative to the pristine monolayer, ensuring that comparisons emphasize the relative energetic hierarchy rather than strict magnitudes.

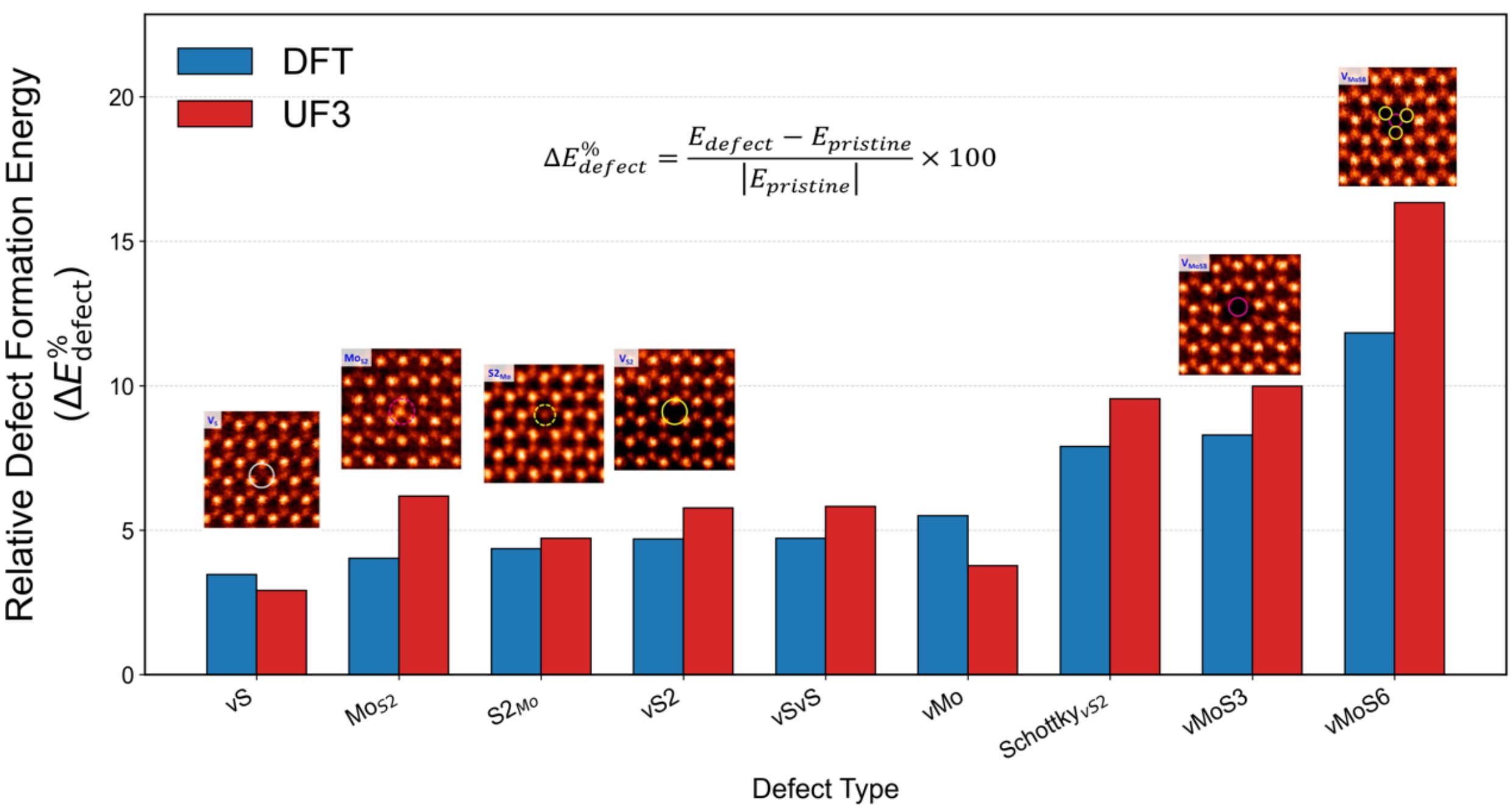

**Fig. 7. Relative defect formation energies in 1H-$MoS_2$ computed with UF3 and DFT.** Insets show experimental images from atomic resolution annular dark field (ADF) imaging.[65]

The ability of UF3 to capture the general stability trend of monolayer defects is further emphasized in Figure 8, which compares UF3 and DFT formation energies for a 75-atom 1H-$MoS_2$ supercell. The correlation is strong ($R^2 = 0.91$), indicating that UF3 reproduces the hierarchy of defects with good fidelity: sulfur vacancies lie at the low-energy end, antisites and the Mo vacancy occupy an intermediate regime, and multi-vacancy complexes ($V_{MoS_3}$, $V_{MoS_6}$, $Schottky_{vS_2}$) are consistently the most costly. Quantitatively, UF3 shows some systematic shifts: it overestimates the energies of highly disrupted environments like $V_{MoS_6}$ and $V_{MoS_3}$ and raises the $Mo_{S_2}$ antisite relative to $S_{2Mo}$. In contrast, it underestimates the Mo vacancy ($V_{Mo}$). However, the trend and ordering that govern defect stability are largely preserved, which is important for predictive simulations of defect environments such as those seen during the kinetic processes of epitaxy.

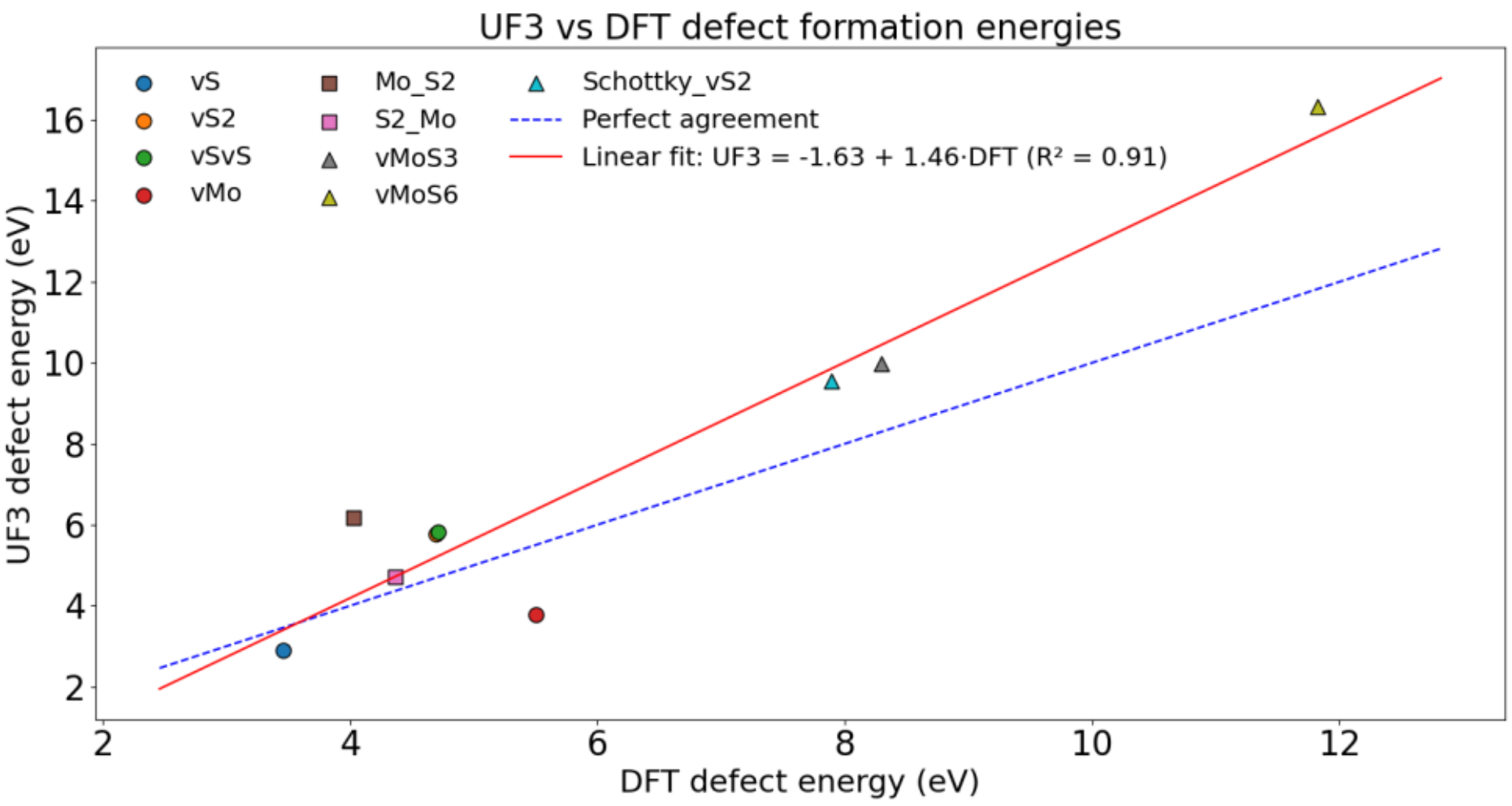

**Fig. 8. Relative defect formation energies parity plot: UF3 vs DFT**

To assess the UF3 potential's ability to capture the stability of $MoS_2$ edges, the edge energy was computed as:

$$\gamma^{edge} = \frac{E_{ribbon} - N e_{monolayer}}{2L},$$

where $E_{ribbon}$ is the energy of the nanoribbon with $N$ total atoms with two exposed edges each of length $L$ and $e_{monolayer}$ is the per-atom energy of the pristine monolayer.

Because UF3 exhibits a systematic cohesive energy offset relative to DFT, comparisons focused only on reproducing the ordering and ratio of edge free energies rather than absolute values. Indeed, the absolute edge free energies predicted by UF3 are nearly uniformly reduced — by approximately 60% — consistent with the global cohesive-energy offset of the potential. Critically, UF3 preserves the energetic ordering between zigzag (ZZ) and armchair (AC) terminations, reproducing their difference and ratio within 5% of DFT as shown in **Table III**. This demonstrates that, despite cohesive energy offsets, UF3 captures the *relative* formation energies of $MoS_2$ edges with excellent fidelity. The free energies of edges determine the equilibrium shape of 2D material flakes and are a plausible thermodynamic driving force in the structure of epitaxially grown 2D material overlayers, with domain morphology favoring to expose low-energy edges[66,67].

**Table III. Edge Energetics in Monolayer $MoS_2$: DFT vs UF3.**

| Property | DFT | UF3 | Error |
|---|---|---|---|
| $\boldsymbol{\gamma}_{ZZ}^{edge}$ (eV/nm) | 1.103 | 0.467 | |
| $\boldsymbol{\gamma}_{AC}^{edge}$ (eV/nm) | 1.143 | 0.509 | |
| $\Delta\boldsymbol{\gamma}^{\boldsymbol{edge}} = \boldsymbol{\gamma}_{ZZ}^{edge} - \boldsymbol{\gamma}_{AC}^{edge}$ (eV/nm) | -0.040 | -0.042 | 5.0 (%) |
| $\boldsymbol{\gamma}_{ZZ}^{edge} / \boldsymbol{\gamma}_{AC}^{edge}$ | 0.965 | 0.917 | - 4.9 (%) |

### *3.5 Thermal Stability*

To assess high-temperature behavior of 2H-$MoS_2$ with the UF3 potential, we evaluate temperature-dependent radial distribution functions and characteristic temperatures associated with surface desorption ($T_{\text{desorb}}^{\text{surface}}$), loss of structural fidelity in multilayer slabs $\left(T_{limit}^{slab}\right)$, loss of structural fidelity in bulk single crystals ($T_{limit}^{bulk}$), and the equilibrium melting temperature ($T_m$) obtained from the two-phase (coexistence) method[68,69]. These quantities are summarized in **Table IV,** with methodological details provided in the Supplemental Material (Figs. S2 – S4 and Supplementary Note 1).

**Table IV. Thermal Stability of 2H-$MoS_2$ predicted by UF3**

| Symbol | Definition | Temperature |
|---|---|---|
| $T_{\text{desorb}}^{\text{surface}}$ | Temperature at which atoms at the (0001) surface begin to thermally desorb into vacuum. | ~2150 K |
| $T_{limit}^{slab}$ | Temperature at which the structure of a (0001) multilayer slab breaks down during heating. | ~ 2550 K |
| $T_{limit}^{bulk}$ | Temperature at which the structure of a bulk single crystal breaks down during heating. | ~ 2700 K |
| $T_m$ | Equilibrium melting temperature as obtained from the two-phase (coexistence) method[68,69]. | ~1650–1700 K |

Temperature-dependent radial distribution functions indicate that the structure of 2H-$MoS_2$ remains preserved until approximately 2700 K (SM, Fig. S2). Complementary simulations show that sulfur atoms begin to thermally desorb from exposed (0001) surfaces of multilayer slabs near ~2150 K, with structural collapse occurring near ~2550 K (SM, Fig. S3). The high-temperature structural stability limits of bulk and multilayer $MoS_2$ predicted by the UF3 MLIP are in reasonable agreement with experimental reports of $MoS_2$ heating without melting at temperatures up to ~2073 ± 20 K, with theoretical estimates suggesting an intrinsic melting limit near ~2648 K[70].

The equilibrium melting temperature of 2H-$MoS_2$ predicted by the UF3 MLIP (~1650 – 1700 K) from coexistence simulations is also physically reasonable. Experimental studies on molybdenite report thermal decomposition in the range ~ 1423 – 1773 K at low pressures (~20 Pa)[71], with thermodynamic analyses predicting similar transition temperatures between ~1573–1673 K under comparable conditions[72]. Ab initio molecular dynamics simulations performed on much smaller systems (≤250 atoms) have predicted melting temperatures in the range ~2150 – 2270 K; however, these estimates may be influenced by finite-size effects, as noted by the author[73]. Most importantly, the equilibrium melting temperature predicted by UF3 lies comfortably above the temperatures employed in the epitaxial growth simulations (~900–1450 K, see Section 3.7); thus, the simulated dynamics occur entirely within the solid-state regime.

### *3.6 Computational Speed*

Simulating epitaxial growth requires an interatomic potential that is both highly accurate and computationally efficient as the process is inherently non-equilibrium, spans large system sizes, and must be simulated over extended timescales often at elevated temperatures. Figure 9 compares the computational performance of several $MoS_2$ interatomic potentials, evaluated using a 54,000-atom supercell over 40,000 MD timesteps (NVE ensemble). Benchmarks were conducted using 6 *AMD EPYC 75F3* CPU cores per job. Classical empirical potentials such as REBO[19], SW-Jiang[20], and SW-FM[20] were the fastest, with total wall times of approximately 0.5 – 1 hour. ReaxFF[25] and SNAP[26] exhibited significantly higher computational cost (~7.6 h and ~15.2 h, respectively). The UF3 MLIP achieved near-classical efficiency with a wall time of ~1.7 h, making it well suited for large-scale, long-timescale epitaxial simulations where both accuracy and speed are essential.

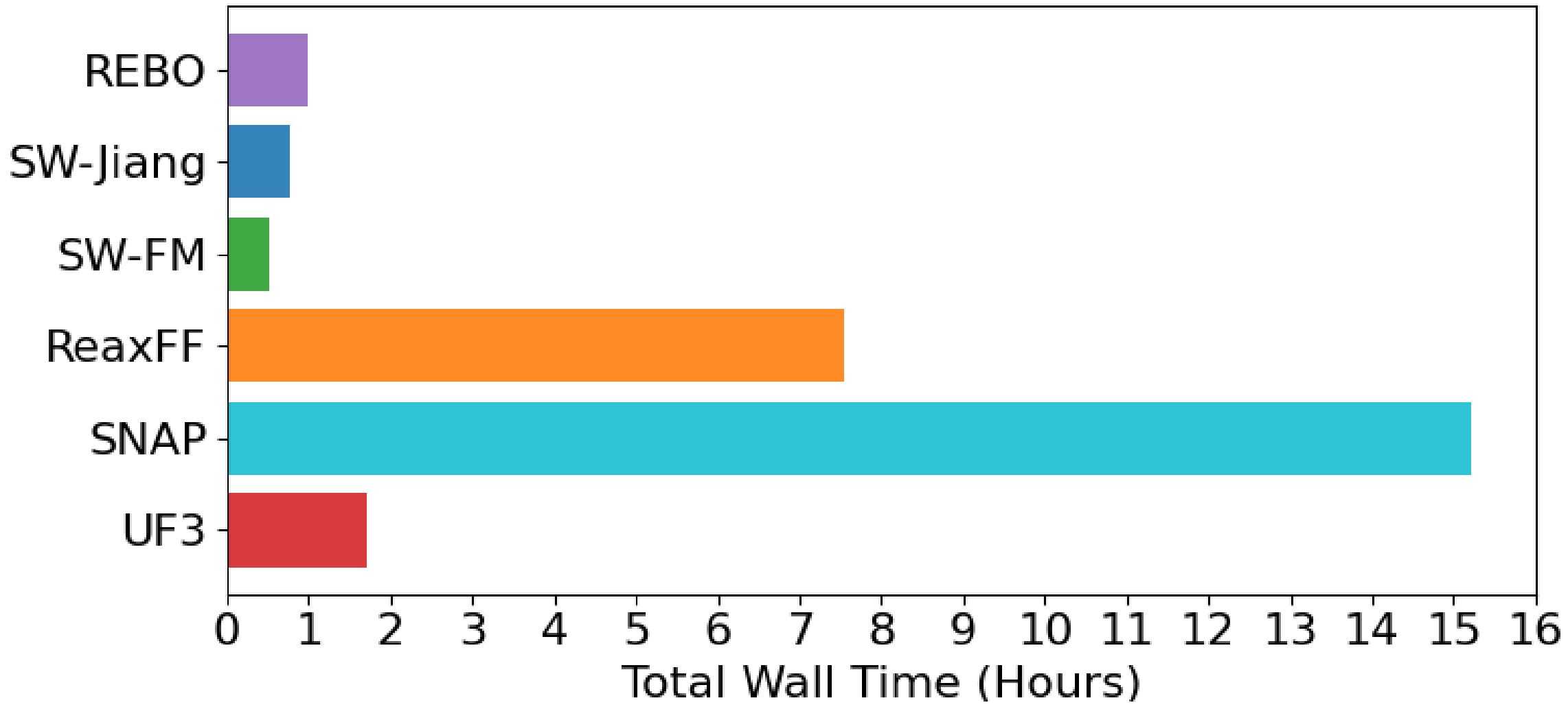

**Fig. 9. Speed comparison of various $MoS_2$ interatomic potentials.** Wall times are reported for a 54,000-atom supercell simulated for 40,000 MD timesteps (NVE ensemble). REBO, SW-Jiang, SW-FM are the fastest, exhibiting wall times between ~ 0.5 and 1 h; ReaxFF and SNAP incur higher computational cost (~7.6 h and ~15.2 h, respectively), while the UF3 ML potential achieves near-classical computational efficiency (~1.7 h).

We note that both MACE-MP-0[59] and CHGNet[60] universal potentials were unable to perform the 54,000-atom benchmark due to out-of-memory errors, consistent with documented performance limits that report memory bottlenecks emerging between 6,250-14,700 atoms depending on hardware[74]. Therefore, to provide a direct comparison between these universal MLIPs and UF3, we perform separate benchmark simulations on a 6,000-atom supercell for 1,000 MD steps using 32 *AMD EPYC 9655P* CPU cores and 256 GB of RAM per job. The total wall times for UF3, MACE-MP-0, and CHGNet were 3.1 s, 1,916.9 s (~32 minutes), and 3,782.7 s (~63 minutes), respectively, corresponding to speedups of over two orders of magnitude for UF3 relative to MACE-MP-0 and CHGNet.

### *3.7 Epitaxial Growth*

To demonstrate the desired capability of the developed potential, we simulated the homoepitaxial growth of $MoS_2$ layers. Epitaxial growth is a challenging process to simulate as it requires a potential that is physically robust and transferable to a wide-range of atomic environments encountered during the highly non-equilibrium process. To the best of our knowledge, existing potentials for $MoS_2$ have yet to demonstrate its epitaxial growth processes.

During simulation, the substrate temperature was maintained using a Langevin thermostat, while the growing epitaxial layers evolved under the NVE ensemble. A reflective boundary was placed above the surface to prevent atoms floating out of the simulation cell. During growth, Mo and S atoms were randomly injected above the substrate at a 1:2 stoichiometric ratio with a small net downward velocity. Multiple growth simulations were run with growth temperatures ranging from 900 K to 1450 K, with deposition rates ranging from ~0.05 to ~0.006 monolayers per nanosecond. The UF3-$MoS_2$ GitHub associated with this work provides LAMMPS input files for the simulation of epitaxial growth.

Figure 10 shows the orthographic side view of $MoS_2$ homoepitaxial growth obtained using the UF3 potential. The 12-layer 26,244-atom 2H-$MoS_2$ substrate is ~ 8.5 × 7.4 nm² in-plane and 7.2 nm thick. The orange dashed line marks the onset of the epilayer region, separating the deposited overlayer from the underlying substrate. The growth direction is perpendicular to the basal plane (0001). After 55 nanoseconds of simulated deposition at 900 K, the overlayer reached a thickness of approximately 2.2 monolayers. The structure preserves the characteristic van der Waals gap between successive layers, demonstrating that the potential maintains weak interlayer interactions while enabling epilayer crystallization. These results confirm that the UF3 potential effectively captures both the structural anisotropy and kinetic stability of $MoS_2$ under extended nonequilibrium growth conditions.

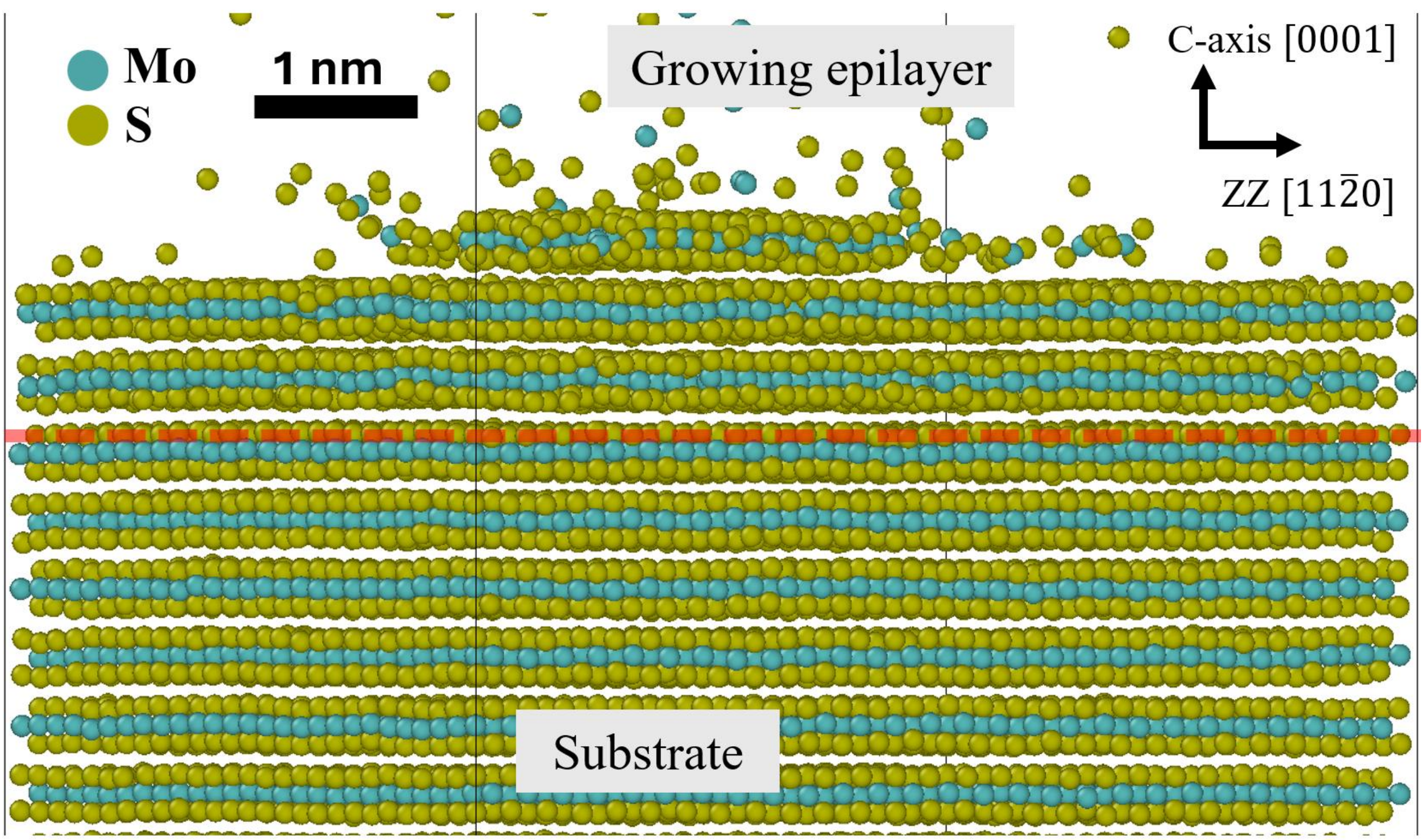

**Fig. 10. Orthographic side view of the atomic structure of homoepitaxially grown $MoS_2$ layers at 900 K obtained via MD simulation.** The 2.2 monolayer thick epilayer maintains the characteristic van der Waals gap between successive layers, with the red dashed line marking the onset of the epilayer region (the interface between the substrate and grown epilayer). Visualized using OVITO[75].

During epitaxial growth, the overlayer observes the formation of a characteristic triangular $MoS_2$ flake. The initial nucleation seed (Fig. 11a) and the resulting epitaxial domain (Fig. 11b, c) exhibit triangular morphology. The substrate consists of 104,976 atoms and is approximately 17.0

× 14.7 nm² in-plane and 8 nm thick; the growth rate was ~0.006 ML/ns and the growth temperature was 1450 K. The triangular domain in Fig. 11c has side lengths of ~16 nm and internal angles of ~ 60°. The edges of the S-terminated triangular seed are oriented along the ⟨1$\bar{1}$$\bar{2}$0⟩ ZZ direction, reflecting edge anisotropy of $MoS_2$, as shown in **Table III**, where zigzag edges are energetically favored over armchair terminations ($\gamma_{ZZ}^{edge} < \gamma_{AC}^{edge}$). By exposing the low energy zigzag edges, the growing layer reduces its free energy relative to other possible morphologies. As growth proceeds, the triangular seed grows laterally, maintaining its aspect ratio and forming a progressively larger domain. Triangular grain formation observed during growth indicates that the UF3 potential qualitatively captures the essential thermodynamic and kinetic factors governing nucleation, edge stability, and shape evolution during epitaxy. Overall, the simulated morphology appears physically consistent with experimentally observed triangular $MoS_2$ seeds and domains (Fig. 11d), demonstrating the potential's ability to reproduce key features of epitaxial growth behavior. The occurrence of triangular domains in epitaxially grown $MoS_2$ films has been widely reported experimentally[8,76–82].

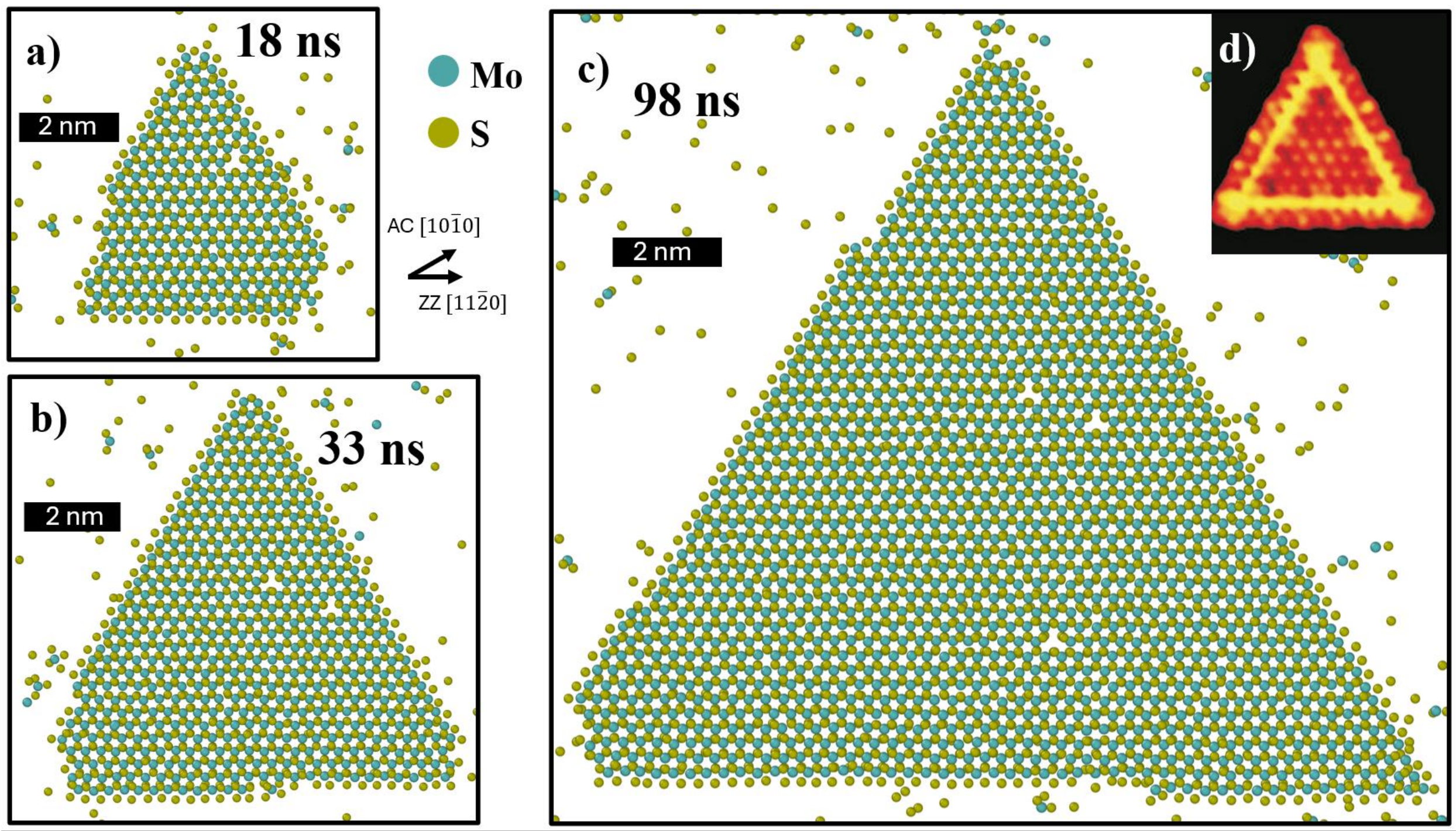

**Fig. 11. Top view of the atomic structure of a homoepitaxially grown $MoS_2$ triangular domain at 1450 K obtained via MD simulation at (a) 18 ns, (b) 33 ns, and (c) 98 ns.** Visualized using OVITO[75]; for clarity, only the epilayer region is shown. **(d) Experimental Scanning Tunneling Microscope (STM) image of triangular $MoS_2$ nanocluster synthesized on Au(111)[82].**

We observe a strong kinetic dependence on the quality of the grown epilayers, with higher temperatures and slower deposition rates producing fewer defects and more well-defined triangular domains, as shown in Fig. 12. The defect densities of the overlayers shown Fig 12 (a), (b), and (c) are approximately 2.1, 0.35, and 0.094 $nm^{-2}$, respectively. Importantly, however, regardless of the kinetic parameters, the formation of vdW gaps between successive layers and triangular domain nucleation are observed. An inherent limitation of MD is the restricted timescale accessible with the approximately femtosecond timestep required to resolve atomic vibrations, which limits the sampling of kinetically activated processes such as adatom diffusion and epilayer formation at experimentally relevant frequencies[83]. Indeed, the deposition rates employed are necessarily many orders of magnitude faster (~$10^5$–$10^6$) than in experiments and further reducing the deposition rate to approach experimental timescales is computationally prohibitive. To compensate for the reduced timescale of MD, we increase the growth temperature relative to experiment to enhance surface diffusion and promote adatom incorporation (1200–1450 K in simulation vs. ~ 823–1273 K in experiment). The characteristic size of the triangular domain grown in simulations being proportional to the simulation cell-size (~8.5 x 7.4 $nm^2$ in-plane) is due to the nanoscale limitations of MD. Of course, under experimentally realistic deposition rates and substrate sizes, the epilayers would be expected to be large in area and exhibit near-perfect crystalline quality.

It is worth noting that classical 2D nucleation theory predicts that the free energy barrier for nucleating a new atomic row along an edge scales with the square of the edge energy[84,85]. Accordingly, the underestimation of edge energies by UF3 relative to DFT (**Table III**) is expected to have the *fortuitous* result of accelerating the kinetic processes at the edges of $MoS_2$ flakes in simulation, further compensating for the limited timescales accessible to MD and increasing the sampling of structures expected on much longer, experimentally relevant timescales. Moreover, sampling of the potential energy surface of S and Mo adatoms across the basal (0001) surface of 2H-$MoS_2$ probes maximum energy differences of ~102 meV (DFT: ~106 meV) and ~190 meV (DFT: ~306 meV), respectively (SM, Fig. S5a). These values are comparable to the thermal energy scale at the growth temperatures considered here (900–1450 K, $k_B T \approx 78$–$125$ meV), indicating that adatoms, particularly sulfur adatoms, are highly mobile during the epitaxial growth simulations. Additionally, sampling the energy of an S adatom along a constrained trajectory descending a zigzag edge suggests that UF3 may underestimate energy barriers associated with adatom descent relative to DFT (SM, Fig. S5b). This underestimation represents a limitation of the UF3 model and may affect quantitative interpretations of the simulated growth dynamics despite the preservation of experimentally observed qualitative features.

We reiterate that morphologies exposing ZZ edges are observed across multiple independent runs and remain robust to variations in growth temperature and deposition rate, as shown in Fig. 12. Furthermore, on a much larger >400,000-atom substrate (~34.0 × 29.5 × 7.5 $nm^3$), ZZ-edge-terminated clusters with facets separated by ~60° remain dominant, forming multiple triangular-like and hexagonal-like domains (see SM, Fig. S6). The van der Waals epitaxial growth of $MoS_2$ hexagonal domains has also been observed experimentally[86].

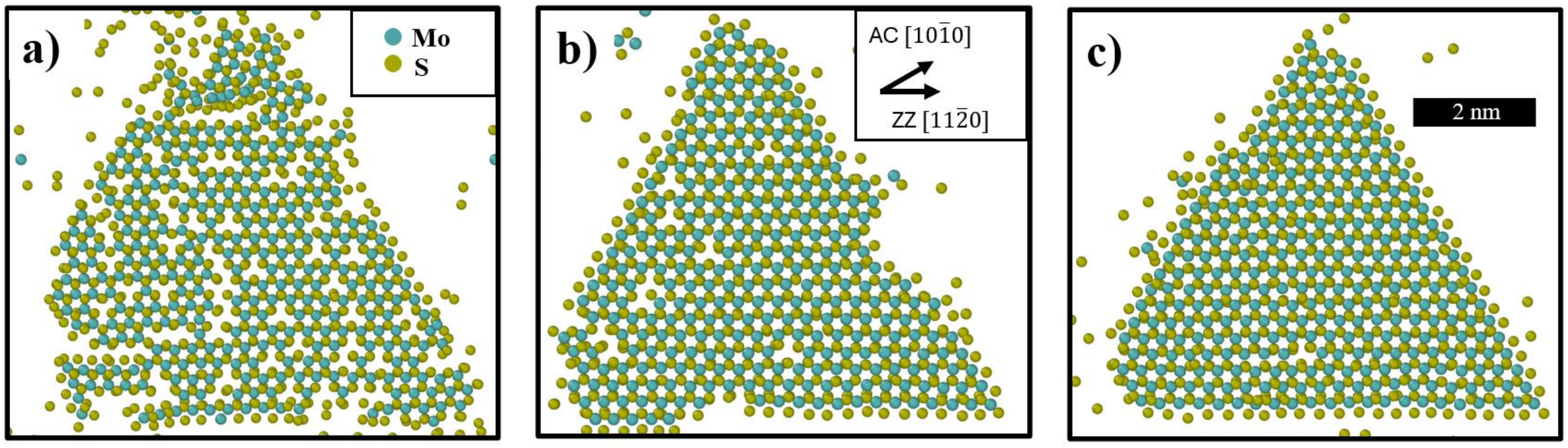

**Fig. 12. Top views of the atomic structures of various homoepitaxially grown $MoS_2$ epilayers at (a) 900 K and ~0.05 ML/ns, (b) 1200 K and ~0.02 ML/ns, and (c) 1450 K and ~0.006 ML/ns, obtained via MD simulations, demonstrating a kinetic dependence on the quality of the grown structures. The corresponding defect densities of the triangular domains are (a) ~2.1 $nm^{-2}$, (b) ~0.35 $nm^{-2}$, and (c) ~0.094 $nm^{-2}$.**

## IV. Summary and Conclusion

In this work, an ultra-fast machine-learned interatomic potential (UF3 MLIP) for $MoS_2$ was developed to enable predictive simulation of the homoepitaxial growth processes. The potential was trained on a comprehensive dataset encompassing equilibrium and non-equilibrium configurations including those obtained from genetic algorithms, aiMD trajectories, and distorted, strained, surface, and low-dimensional structures. The hyperparameters were optimized using the UF3Tools software package[38].

The UF3 potential reproduces lattice parameters and relative phase stability of $MoS_2$ polymorphs (1H, 2H, 3R, 1T′) with near-DFT accuracy, while maintaining only twice the computational cost of the fastest empirical potentials. The UF3 potential captures long-range van der Waals binding necessary for layered structures, reproducing basal-plane surface energies within ~2 % of DFT values for 2H-$MoS_2$. Phonon spectra are reproduced with sub-0.5 THz RMS deviation across acoustic and optical branches, and the highly anisotropic elastic tensor is reasonably captured. Critically, the UF3 potential captures defect and edge energetics with high fidelity: the hierarchy of formation energies for ten monolayer $MoS_2$ defect structures is reproduced with strong correlation to DFT ($R^2 = 0.91$) and the ratio and difference between the free energies of zigzag and armchair edges is reproduced within 5% of DFT.

Most importantly, in molecular-dynamics simulations, the potential reproduces two experimentally observed features of $MoS_2$ epitaxial growth: (i) formation of van der Waals gaps between successive epilayers and (ii) zigzag-edge-terminated triangular domain formation. To our knowledge, this represents the first experimentally consistent MD simulation of $MoS_2$ epitaxial growth. The developed UF3 MLIP provides a critical step toward the modelling and prediction of the epitaxial growth of $MoS_2$ layers. Future work will focus on (1) a comprehensive investigation of the epitaxial growth dynamics of $MoS_2$ layers, including growth on mesoscopic substrates (lateral dimensions $\gtrsim 0.1$ μm), and (2) the development of a $MoS_2$/substrate potential to enable large-scale heteroepitaxial growth simulations.

*Acknowledgements*

This work is based on research supported by the US National Science Foundation under Award Number CMMI- 2349160. The work of Phillpot is supported by DMR 2121895. The work of Chen is also partially supported by DMR 2121895. The computer simulations are funded by the Advanced Cyberinfrastructure Coordination Ecosystem: Services & Support (ACCESS) allocation TG-DMR190008. We thank Jason Gibson, Salil Bavdekar, Michael MacIsaac, Stephen Xie, and Matthias Rupp for insightful discussions related to this project.

*Data Availability*

Data will be made available on reasonable request. The UF3 potential file, input scripts for epitaxial growth simulations, and the Supplemental Material are available at https://github.com/uf-chenlab/UF3-MoS2. The UF3Tools code can be found at: https://github.com/uf-chenlab/UF3Tools.